\documentclass[12pt]{iopart}
\usepackage{amssymb,amsfonts,textcomp}

\newcommand\textstyleHTML[1]{\textup{#1}}

\begin{document}
\title[Maximum entropy principle in NSO]{Maximum entropy principle
and the form of source in non-equilibrium statistical operator method}

\author{V Ryazanov}

\address{Institute for Nuclear Research,
pr.Nauki, 47, Kiev, Ukraine}

\ead{vryazan@kinr.kiev.ua}

\begin{abstract}
It is supposed that the exponential multiplier in the method of the
non-equilibrium statistical operator (Zubarev`s approach) can be
considered as a distribution density of the past lifetime of the
system, and can be replaced by an arbitrary distribution function.
To specify this distribution the method of maximum entropy principle as in
[Sch\"onfeldt J-H, Jiminez N, Plastino A R, Plastino A, Casas M
2007 \textit{Physica A} \textbf{374}~573] is used. The obtained
distribution is close to exponential one. Another approach to
the maximum entropy principle, as in [Van der Straeten E and Beck
C 2008 Phys. Rev. E \textbf{78} 051101], except exponential
distributions yields power-like,
log-normal distributions, as well as distributions of other kind and
transitions between them.
\end{abstract}


\maketitle

\section{Introduction}

Among possible approaches to the description of non-equilibrium
systems the \textit{Non-equilibrium Statistical Operator Method}
(NESOM) especially demonstrated its efficiency \cite{zub74,zub80,zub95}.
NESOM provides a very promising technique that implies in
a far-reaching generalization the statistical methods developed
by Boltzmann and Gibbs. NESOM was initially built on intuitive and
heuristic arguments, apparently it can be incorporated within an
interesting approach to the rationalization of statistical
mechanics, as contained in the maximization of (informational
statistical) entropy (MaxEnt for short) and Bayesian methods.
NESOM appears as a very powerful, concise, based on sound
principles, and elegant formalism of a broad scope to deal with
systems arbitrarily far from equilibrium. The non-equilibrium
statistical operator (NSO) introduced in \cite{zub74,zub80,zub95}
has a form
\begin{equation}
\ln\varrho(t)= \int_{0}^{\infty}p_{q zub}(u)\ln\varrho_{q}(t-u,
-u)du,\quad
\ln\varrho_{q}(t, 0)=-\Phi(t)-\sum_{n}F_{n}(t)P_{n};\\
\label{zu}
\end{equation}
$$
\ln\varrho_{q}(t, t_{1})=\e^{\left\{-t_{1}H/i\hbar\right\}}\ln\varrho_{q}(t,
0)\e^{\left\{t_{1}H/i\hbar\right\}}; \quad \Phi(t)=\ln Sp
\exp\{-\sum_{n}F_{n}(t)P_{n}\}\,,
$$
where

\begin{equation}
p_{q}(u)=p_{q zub}(u)=\varepsilon \e^{-\varepsilon u}\,, \quad u=t-t_{0}\,, \label{zub}
\end{equation}
$H$ is Hamiltonian, $\ln\rho(t)$ is the logarithm of the NSO,
$\ln\rho_{q}(t_{1}, t_{2})$  is the logarithm of the
quasi-equilibrium (or relevant) distribution; the first time
argument indicates the time dependence of the values of the
thermodynamic parameters $F_{n}$; the second time argument $t_{2}$
in $\ln\rho_{q}(t_{1}, t_{2})$ denotes the time dependence through
the Heizenberg representation for dynamical variables $P_{n}$ on
which $\ln\rho_{q}(t, 0)$ can depend \cite{zub74,zub80,zub95}.

In \cite{zub74,zub80,zub95} $p_{q}(u)=p_{qzub}(u)=\varepsilon
\exp\{-\varepsilon u\}$; after the thermodynamic limiting
transition $N\rightarrow\infty$, $V\rightarrow\infty$,
$N/V=const$, $\varepsilon\rightarrow 0$. From the complete group
of solutions of Liouville equation (symmetric in time) the subset
of retarded ''unilateral'' in time solutions is selected by means
of introducing a source $K$ in the Liouville equation for
$\ln\rho(t)$ ($L$ is Liouville operator;
$iL=-\{H,\varrho\}=\sum_{k}{\displaystyle\left[\frac{\partial H}{\partial
p_{k}}\frac{\partial\varrho}{\partial q_{k}}-\frac{\partial
H}{\partial q_{k}}\frac{\partial\varrho}{\partial p_{k}}\right]}$;
$p_{k}$, and $q_{k}$ are pulses and coordinates of particles;
$\{\dots\}$ are Poisson brackets)

\begin{equation}
\frac{\partial\ln \rho(t)}{\partial
t}+iL\ln\rho(t)=-\varepsilon(\ln\rho(t)-\ln\rho_{q}(t,
0))=K_{zub}. \label{sour}
\end{equation}

In \cite{Rop,Der} a convenient redefinition of the source term is
proposed. Although infinitesimally small, the source term
introduced by Zubarev into the Liouville equation is shown to
influence the macroscopic behaviour of the system in the sense
that the corresponding evolution equations do not coincide exactly
with those obtained from an initial-value problem which
corresponds to a definite experimental situation and a physical
set of macroobservables.

In \cite{alg,ram} it was noted that in place of the function
$p_{q}(u)=p_{q zub}(u)=\varepsilon \e^{-\varepsilon u}$ in
(\ref{zu}) arbitrary (but having certain properties \cite{luz}) weigth
functions $w(t, t_{0})$ can be used. Zubarev's nonequilibrium
statistical operator does satisfy Liouville equation, but it must
be borne in mind that the group of its solutions is composed
of two subsets, one corresponding to the retarded
and second one to the advanced solutions. The presence
of the weight function $w(t, t_{0})$ (Abel's kernel (\ref{zub}) in
Zubarev's approach) in the time-smoothing or quasi-average
procedure that has been introduced selects the subset of retarded
solutions from the total group of solutions of the Liouville equation.
This consideration is related to the question: how to obtain an
irreversible behavior in the evolution of the macroscopic state of
the system? In the MaxEnt-NESOM approach the irreversibility is incorporated from
the outset using an {\it ad hoc} non-mechanical hypothesis.
MaxEnt-NESOM yields information on the macrostate of the system
at time $t$, when a measurement is performed, including the
evolutionary history (in the interval from the initial time of
preparation $t_{0}$ up to time $t$) by which the system came into
that state (which introduced a generalization of Kirkwood`s \
time-smoothing formalism \cite{kir}). Functions $w(t, t_{0})$
are typically kernels \cite{zub74,zub80,zub95,luz} that appear in
the mathematical theory of convergence of integrals. In
\cite{ry01,ry07,ry09} other interpretation of the functions $w(t,
t_{0})$, denoted as $p_{q}(u)$, is given. With the change of
function $w(t, t_{0})=p_{q}(u)$ the form of source
(\ref{sour}) in the Liouville equation also changes. For an arbitrary function
$p_{q}(u)$ it looks like (\ref{so}).

In \cite{ry01,ry07,ry09} it was noted that the function
$p_{q}(u)=p_{q zub}(u)=\varepsilon \e^{-\varepsilon u}$ in NESOM
\cite{zub74,zub80,zub95} for the non-equilibrium distribution function
can be interpreted as the exponential probability distribution of
the lifetime $\Gamma$ of a system. $\Gamma$ is a random variable
of lifetime (time span) from the moment $t_{0}$ of its birth till
the current moment $t$; $\varepsilon^{-1}=\langle t-t_{0}\rangle
=\langle\Gamma\rangle$, where $\langle\Gamma\rangle =\int u
p_{q}(u)du$ is the average lifetime of the system. This time
period can be called the time period of getting information about
system from its past. Instead of the exponential distribution
$p_{q zub}(u)$ (\ref{zub}) in (\ref{zu}) used in
\cite{zub74,zub80,zub95} any other sample distribution $p_{q}(u)$
could be taken; integration by parts in time is performed at $\int
p_{q}(u)du_{|u=0}=-1$; $\int p_{q}(u)du_{|u\rightarrow\infty}=0$.
If $p_{q}(u)=p_{q zub}(u)=\varepsilon e^{-\varepsilon u}$ by
(\ref{zub}), $\varepsilon=1/\langle\Gamma\rangle $ the expression
for NSO passes in (\ref{zu}) from \cite{zub74,zub80,zub95}.

The same interpretation of the distribution  $p_{q}(u)$ is given in
\cite{zub80}, where this value is understood as the distribution
of the initial moment of time $t_{0}$. Since the random (past) lifetime
is equal to $\Gamma=u=t-t_{0}$, the distribution of the past
lifetime ${u}$ coincides with the distribution of the initial time values
$t_{0}$. The moment $t_{0}$ will be the moment of the first passage in
the inverse time, if the moment $t$ is taken as initial. In
\cite{zub80} the uniform distribution for an initial moment
$t_{0}$ is chosen, which after the transition from Abel integration
to Ces\`aro integration passes to the exponential distribution
$p_{q}(u)=p_{q zub}(u)=\varepsilon \e^{-\varepsilon u}$. Such
distribution serves as the limiting distribution of the lifetime
\cite{str}, the first-passage time of a certain level. In the general case
it is possible to choose a lot of functions for the obvious type
of distribution $p_{q}(u)$, which was noted in
\cite{ry01,ry07,ry09}.

In \cite{str} the lifetimes of the system are introduced as
random moments of the first-passage time till the moment when a
random process describing system reaches a certain limit, for
example, a zero value. In \cite{str} approximate exponential
expressions for the probability density function (with a single
parameter) and probability distribution of lifetime are obtained,
and the accuracy of these expressions is estimated.

In \cite{mor} it was noted, that the role of the form of the source
term in the Liouville equation in the NSO method has never been
investigated. In \cite{fel} it is stated that the exponential
distribution is the only one which possesses the Markovian
property of the absence of afteraction, that is whatever is the
actual age of a system, the remaining time does not depend on the
past and has the same distribution as the lifetime itself. It is
known \cite{zub74,zub80,zub95} that the Liouville equation for NSO
contains the source
$K=K_{zub}=-\varepsilon[\ln\rho(t)-\ln\rho_{q}(t, 0)]$
(\ref{sour}) which becomes vanishingly small after taking the
thermodynamic limit and setting $\varepsilon\rightarrow 0$, which
in the spirit of \cite{ry01} corresponds to the infinitely large
lifetime value of an infinitely large system. For a system with
finite size this source is not equal to zero. In \cite{luz} this
term enters the modified Liouville operator and coincides with the
form of Liouville equation suggested by Prigogine \cite{pri} (the
Boltzmann-Prigogine symmetry), when the irreversibility is
introduced in the theory at the microscopic level.

In \cite{ry01} a new interpretation of the method of the NSO is given,
in which the operation of taking the invariant part \cite{zub74,zub80}
or the use of an auxiliary ''weight function'' (in the terminology of
\cite{alg,ram,luz}) in NSO are treated as averaging the
quasi-equilibrium statistical operator over the distribution of past
lifetime of a system. This approach agrees with
the approach of
the general theory of random processes, the renewal theory,
and also with the conception of Zubarev work \cite{zub80} where the
NSO is conceived as some averaging over the initial moment of time.

The statistical operator depends on the information-gathering
interval $(t_{0}, t)$, but it must be borne in mind that this is
the formal point consisting in that (as Kirkwood pointed out) that the
description to be built must contain all the previous history in
the development of the macrostate of the system. In
\cite{alg,ram,luz} several basic steps for the construction of the
NESOM formalism are indicated: a third basic step has just been
introduced, namely, the inclusion of the past history (other terms used are
retro-effects or historicity) of the macrostate
of the dissipative system. A fourth basic step needs now to be
considered, which is a generalization of Kirkwood's time-smoothing
procedure: the one that accounts for the past history and future
dissipative evolution. The time-smoothing procedure introduces a
kind of Prigogine's dynamical condition for dissipativity. The
procedure introduces a kind of evanescent history as the system
macrostate evolves toward future from the initial condition at
time $t_{0}$. The function $w(t; t_{0})$ \cite{alg,ram,luz}
introduces the time-smoothing procedure. In principle,
any kernel  provided by the
mathematical theory of convergence
of trigonometrical series and  transform integrals provides is
acceptable for these purposes.
Kirkwood, Green, Mori \cite{kir,gre,mori} and others have chosen
what in mathematical terminology is Fej\`{e}r (or Ces\`{a}ro-1)
kernel. Meanwhile Zubarev introduced the one consisting in Abel's
kernel for $w$ in Eq. (\ref{zu}) - which apparently appears to be
the best choice, either mathematically but mostly physically:
that is, taking $w(t; t_{0}) = \varepsilon
\e^{\varepsilon(t_{0}-t)}$, where $\varepsilon$ is a positive
infinitesimal value which tends to zero after the calculation of averages
has been performed, and with $t_{0}$ going to $- \infty $.
Therefore a process with fading memory is introduced.
 In Zubarev's approach
this fading process occurs in an adiabatic-like form towards the
remote past: as time evolves memory decays exponentially with
lifetime $\varepsilon^{-1}$ \cite{luz}. The approach suggested in
\cite{ry01,ry07,ry09} and in the present work enables to use a family
of functions $w(t,t_{0})=p_{q}(u)$ and makes clear both their
physical sense and those physical situations in which one or
another function $w(t,t_{0})=p_{q}(u)$ can be used.

Besides the Zubarev's form of NSO \cite{zub74,zub80,zub95}, the NSO formulation in
the Green-Mori form \cite{gre,mori} is known, where one assumes the
auxiliary weight function \cite{alg,ram,luz} to be equal to $W(t,
t_{0})=1-(t-t_{0})/\tau$; $w(t, t_{0})=dW(t,
t_{0})/dt_{0}=1/\tau$; $\tau=t-t_{0}$. After averaging one sets
$\tau\rightarrow\infty$. This choice at $p_{q}(u=t-t_{0})=w(t,
t_{0})$ coincides with the uniform lifetime distribution. The
source in the Liouville equation takes the form
$K=\ln\rho_{q}/\tau$. In \cite{zub74} this form of NSO is compared
to the Zubarev's form. One could name many examples of explicitely
setting the function $p_{q}(u)$. Each and every definition implies some
specific form of the source term $K$ in the Liouville equation,
some specific form of the modified Liouville operator and NSO
\cite{ry01,ry07,ry09}. Thus the whole family of NSO is defined.

It is possible to make different assumptions about the form of the
function of  $p_{q}(u)$, getting different expressions for the
source in the Liouville equation and for non-equilibrium
characteristics of the system. It is possible to show \cite{ry09}
that certain choices of the function of $p_{q}(u)$ result in the
changes in non-equilibrium characteristics in the limit of
infinitely large average lifetimes as well. In
\cite{ry01,ry07,ry09} an analogy is traced to the passage to the
thermodynamic limit of systems of infinite size. So explicit form
of the function $p_{q}(u)$ is important for describing
non-equilibrium systems by the NSO method.

Setting the form of the function $p_{q}(u)$ reflects not only the
internal properties of a system, but also the influence of the
environment on an open system, the particular character of its
interaction with the environment \cite{luz}. In \cite{zub80} a
physical interpretation of the exponential distribution for the
function $p_{q}(u)$ is given: a system evolves freely like an
isolated system governed by the Liouville operator. Besides that
the system undergoes random transitions, and the phase point
representing the system switches from one trajectory to another
one with an exponential probability under the influence of the
"thermostat"; the average intervals between successive push events
increase infinitely. This takes place if the parameter of the
exponential distribution tends to $0$ after the transition to the
thermodynamic limit. Real physical systems have finite sizes. The
exponential distribution describes completely random systems. The
influence of the environment on a system can have organized
character as well, for example, this is the case of systems in a
stationary non-equilibrium state with input and output fluxes.
The character of the interaction with the environment can also
vary; therefore different forms of the function $p_{q}(u)$ can be
used.

The adequate choice of the function $p_{q}(u)$ is important for
correct description of the non-equilibrium properties of
statistical systems. To find the type of function $p_{q}(u)$, it
is necessary to resort to some general principles, such as MaxEnt
principle. In this work  two variants of MaxEnt are used for this
purpose: the one introduced in \cite{sch} for the Liouville
equation with a source (Section 2) and that suggested in
\cite{stBe} for the superstatistics (Section 3).

\section{Maximum entropy principle for Liouville equations with source}

In this paper we apply the maximum entropy principle for the
determination of the function $p_{q}(u)$. The same approach was
applied in \cite{sch} for the evolution equations with source terms.
In \cite{ram,ry01,ry07,ry09} a general form for the source in the
Liouville equation for $\ln\rho(t)$ (\ref{sour}) is obtained. For
our case the source term has the following form

\begin{equation}
K=p_{q}(0)\ln\rho_{q}(t, 0)+\int_{0}^{\infty}\frac{\partial
p_{q}(u)}{\partial u}\ln\rho_{q}(t-u, -u)du. \label{so}
\end{equation}

In \cite{sch} in the Liouville equation
the distribution function \textit{${\rho}$}\textit{(z,t)} is written
in a form

\begin{equation}
\rho(\overrightarrow{z}, t)=Nf_{ME}(\overrightarrow{z},
t)=\frac{N}{Z}\exp\{-\sum_{i=1}^{M}\lambda_{i}A_{i}\},
\label{dis}
\end{equation}
where  $A_{i}(\vec{z})$  are $M$ appropriate quantities that are
functions of the phase space point  ${\vec{z}}$; the quantities
${A_{i}(\vec{z})}$ correspond to the values $P_{i}$ from
(\ref{zu}). The partition function $Z$ is given by

\begin{equation}
Z=\int\exp\{-\sum_{i=1}^{M}\lambda_{i}A_{i}\}d^{N}z. \label{Z}
\end{equation}

The function  ${f_{ME}(\vec{z},t)}$  is normalized to unity:

\begin{equation}
\int f_{ME}(\overrightarrow{z}, t)d^{N}z=1; \label{no}
\end{equation}

\begin{equation}
\int\rho(\overrightarrow{z}, t) d^{N}z=N(t); \quad
\frac{dN}{dt}=\int K d^{N}z;
\label{nab}
\end{equation}
$$
\frac{\partial \rho}{\partial
t}+\overrightarrow{w}\nabla\rho=\frac{d\rho}{dt}=K; \quad \nabla
\overrightarrow{w}=0.
$$

The probability distribution ${f_{ME}(\vec{z},t)}$  is the one
that maximizes the entropy $S[f]$ under the constraints imposed by
normalization and relevant mean values $\langle A_{i}\rangle
=\int A_{i}\rho d^{N}z$ (or $a_{i}=\langle A\rangle_{i}/N)$.
The re-scaled mean values $a_{i}$ and the associated
Lagrange multipliers $\lambda_{i}$ are related by the
Jayne's relations \cite{jay,jay1}

\begin{equation}
\lambda_{i}=\frac{\partial S}{\partial a_{i}},\quad \
a_{i}=\frac{\langle A_{i}\rangle }{N}=-\frac{\partial}{\partial
\lambda_{i}}(\ln Z)\,,
\label{ent}
\end{equation}
$$
S=-\int f\ln f d^{N}z=\ln Z+\sum _{i}\lambda_{i}a_{i}.
$$

If we choose for $\ln\rho(z,t)$ the function

\begin{equation}
\ln\rho=\ln\varrho(t)=
\int_{0}^{\infty}p_{q}(u)\ln\varrho_{q}(t-u,-u)du=
\int_{-\infty}^{t}
p_{q}(t-t_{0})\ln{\rho}_{q}(t_{0},t_{0}-t){dt}_{0} \label{fun}
\end{equation}
from (\ref{zu}), which is included in the Liouville equation
(\ref{sour}) (without integrating over time) with
$p_{qzub}\rightarrow p_{q}, K_{zub}\rightarrow K$, and, following
\cite{zub74}, choose

\begin{equation}
\mathit{\lambda}_{i}=p_{q}(t-t_{0})F_{i}(t_{0}),
\label{la}
\end{equation}
then
\begin{eqnarray}
\ln{\rho}(\vec{z},t)=\int_{-\infty}^{t}
p_{q}(t-t_{0})\ln{\rho}_{q}(t_{0},t_{0}-t){dt}_{0}=
\label{ln} \\
-\int_{-\infty}^{t}(\sum_{i} \lambda_{i}A_{i}+p_{q}(t-t_{0})\ln
Z_{1}){dt}_{0}= \ln f'_{ME}+\ln N \,,\nonumber
\end{eqnarray}
where
$$
\ln f'_{ME}=-\int_{-\infty}^{t}\sum_{i}
\lambda_{i}A_{i}{dt}_{0}-\ln Z_{\lambda};
$$
\begin{equation}
\ln N={\Delta}Z=\ln
Z_{\lambda}-\int_{-\infty}^{t}p_{q}(t-t_{0})\ln
Z_{1}{dt}_{0};\ \label{lnN}
\end{equation}
$$
Z_{\lambda}=\int\exp[-\int_{-\infty}^{t}\sum_{i}
 \lambda_{i}A_{i}{dt}_{0}]d^{N}z,\quad Z_{1}=\int \exp[-\sum_{i}
F_{i}A_{i}]d^{N}z
$$
($F_{i}$ are taken from (\ref{zu})). The values $Z_{\lambda}$ and
$Z_{1}$ in the terminology of \cite{zub74} are related to the partition
functions for a non-equilibrium and relevant statistical
operator accordingly.

In \cite{sch} for the Liouville equation of the kind (\ref{sour}) with
constant sources equation one gets for $d \lambda_{i}/dt$

$$
\frac{d \lambda_{i}}{dt}=(\sum_{i=1}^{M}
C_{ji}{\lambda}_{j})-\frac{1}{N}\frac{\partial}{\partial
a_{i}}\int {K \ln f_{ME}} d^{N}z ,
$$
where the Zubarev-Peletminskiy selection rule \cite{pel,akh,zub74,ram,luz}

\begin{equation}
\vec{{w}}\vec{\nabla }A_{i}=\sum_{j=1}^{M} {C_{ij}}A_{j},\
(i=1,\dots,M),\ \frac{d\vec{z}}{dt}=\vec{w}(\vec{z});\
\frac{\partial \mathit{\rho}}{\partial t}+\vec{\nabla }(\rho
\vec{w})=K \label{pel}
\end{equation}
is used;  $i,j=1,2...$; the $C_{ij}$ are c-numbers. In other
representations the quantities $A_{i}$ can depend on the space
variable, that is, when considering local densities of dynamical
variables, and then the $C_{ij}$ can depend on the space variable as well
or be differential operators.

If more complex shape of the source
(\ref{so}) is considered, the equation for $d\lambda_{i}/dt$ takes on the
form

\begin{eqnarray}
\frac{d\lambda_{i}}{dt}=\left(\sum_{j=1}^{M}
C_{ji}{\lambda}_{j}\right)-\frac{\partial }{\partial
a_{i}}\left(\frac{1}{N}\int {K\ln f_{ME}d^{N}z}\right)-
\label{eq} \\
\sum_{j}{\lambda}_{j}\frac{\partial }{\partial
a_{i}}\left({\displaystyle\frac{1}{N}}\int
 A_{j}K d^{N}z\right)-\ln Z_{1}{\displaystyle\frac{\partial
}{\partial a_{i}}}\left({\displaystyle\frac{\dot{N}}{N}}\right).
\nonumber
\end{eqnarray}

We replace the operators ${\displaystyle\frac{\partial }{\partial a_{i}}}$
and ${\displaystyle\frac{\partial }{\partial \mathit{\lambda}_{i}}}$ taking
into account (\ref{ent})-(\ref{lnN}) by the functional
differentiation of the kind

$$
\frac{\partial}{\partial a_{i}}\rightarrow\frac{\delta}{\delta
a_{i}}=N\frac{\delta}{\delta \langle A_{i}\rangle }, \quad
\frac{\partial}{\partial
\lambda_{i}}\rightarrow\frac{\delta}{\delta \lambda_{i}},
$$
which takes off the integration over time. For example

$$
\frac{\delta}{\delta
\langle A_{i}\rangle }\ln Z_{\lambda}=-p_{q}(t-t_{0})\sum_{k}\frac{\partial
F_{k}}{\partial \langle A_{i}\rangle }\langle A_{k}\rangle ; \quad \frac{\delta
\ln Z_{\lambda}}{\delta \lambda_{i}}=-\langle A_{i}\rangle .
$$
The relations (\ref{dis}-\ref{nab}) and (\ref{ent}) are thus hold.
If to take into account that $\int
\dots\rho_{q}d^{N}z=\int \dots \rho d^{N}z=\langle\dots\rangle$ in the NSO method,
then

\begin{equation}
\frac{\partial }{\partial a_{i}}N=\frac{\partial }{\partial
a_{i}}\left(\frac{\dot{N}}{N}\right)=0 \,. \label{16}
\end{equation}

Let us consider the integrals in the rhs of (\ref{eq}) of the form
$\int K B(z) dz$, $B$ being an arbitrary function of the dynamic
variables $z$, and the source term $K$ taken from (\ref{so}); for Eq.
(\ref{nab}), (\ref{pel}) $K=[p_{q}(0)\ln\rho_{q}(t,
0)+{\displaystyle\int_{-\infty}^{t}\frac{\partial p_{q}(t-t_{0})}{\partial
t}}\ln\rho_{q}(t_{0}, t_{0}-t)dt_{0}]\rho$. Assume that
${p_{q}(u)}$  does not depend on $z$. Integrating by parts and
assuming ${p_{q}(u)_{u\rightarrow \infty }\rightarrow 0}$, we get:

\begin{equation}
\int KB d^{N}z=-\int_{0}^{\infty} p_{q}(u)\frac{d}{du}\langle B
\ln\rho_{q}(t-u, -u)\rangle du; \quad u=t-t_{0} . \label{17}
\end{equation}
Taking into account (\ref{16}) and the
fact that the operation \ ${\displaystyle \frac{\partial }{\partial
a_{i}}\rightarrow
\frac{\mathit{\delta}}{\mathit{{\delta}a}_{i}}}$  eliminates the
integration by time, the equation (\ref{eq}) takes on the form

\begin{equation}
{F_{i}(t_{0})\frac{dp_{q}(t-t_{0})}{dt}=
-p_{q}(t-t_{0})C_{i}-p_{q}(t-t_{0})r_{1}-p_{q}^{2}}(t-t_{0})r_{2},
\label{in}
\end{equation}
where $C_{i}=\sum_{j} C_{ji}F_{j}(t_{0})$,

\begin{equation}
r_{1}=-\frac{\partial }{\partial
\langle A_{i}\rangle }\frac{d}{dt}\langle (\ln{\rho}(t)- \ln
N)\ln{\rho}_{q}(t_{0},t_{0}-t)\rangle ; \label{r1}
\end{equation}

\begin{equation}
r_{2}=-\sum_{j} {F_{j}(t_{0})r_{2j};\; \quad
 r_{2j}=\frac{\partial }{\partial
\langle A_{i}\rangle}\frac{d}{dt}\langle A_{j}(t)
\ln{\rho}_{q}(t_{0},t_{0}-t)}\rangle. \label{r2}
\end{equation}

An unknown function  $p_{q}(u)$ enters the expression
(\ref{r1}) through the terms $\ln\rho(t)$ and $\ln N$. To get rid
of this dependence, we use the averaging theorem. For the
expressions for $\ln{\rho}(t)$ and $\ln N$
 in (\ref{r1}) we take  all terms besides
$p_{q}(u)$ out of the time integration. For each of these function however a
different effective average time value should be used. The remaining integrals
over ${p_{q}(u)}$  are equal to unity. We get:

\begin{equation}
\ln N\simeq \ln Z_{1}(c_{3})-\ln Z_{1}(c_{4}), \label{21}
\end{equation}

\begin{equation}
\ln{\rho}(t)\simeq \ln{\rho}_{q}(c_{1},c_{1}-t)=-(\sum_{m}
F_{m}(c_{1})A_{m}(c_{1}-t)+\ln Z(c_{1})). \label{22}
\end{equation}

Let us make another approximation and change the order of the
operations ${\partial /\partial \langle A_{i}\rangle }$  and $d/dt$  in the
expressions (\ref{r1})-(\ref{r2}). The value
$\int\limits_{t_{0}}^{t}r_{1} dt =D(t)-D(t_{0})$ enters the
expression (\ref{in}), where

\begin{eqnarray}
D(t)=-\frac{\partial }{\partial \langle A_{i}\rangle }\left\langle (\ln{\rho}(t)- \ln
N)\ln{\rho}_{q}(t_{0},t_{0}-t)\right\rangle  \nonumber  \\
= F_{i}(t_{0})[\ln
Z(c_{3})-\ln Z(c_{4})-\ln Z(c_{1})]-  \label{D} \\
F_{i}(c_{1})\ln Z(t_{0})+\sum_{m,n}
\big(\langle A_{m}A_{n}\rangle -\langle A_{m}\rangle \langle A_{n}\rangle \big)
\left[\frac{F_{n}(t_{0})}{\langle A_{m}A_{i}\rangle -\langle A_{m}\rangle \langle A_{i}\rangle }- \right.
\nonumber \\
\left.
-\frac{F_{m}(c_{1})}{\langle A_{i}A_{n}\rangle -\langle A_{i}\rangle \langle A_{n}\rangle }\right]-\sum_{m,n}
F_{m}(c_{1})F_{n}(t_{0}) \sum_{k}
\frac{\langle A_{k}A_{m}A_{n}\rangle -\langle A_{m}A_{n}\rangle \langle A_{k}\rangle }{\langle A_{i}A_{k}\rangle -
\langle A_{i}\rangle \langle A_{k}\rangle } ; \nonumber
\end{eqnarray}

\begin{eqnarray}
r_{2j}=\frac{\partial }{\partial t}\left[\sum_{m}
F_{m}(t_{0}) \sum_{k}
\frac{\langle A_{k}A_{j}A_{m}\rangle -\langle A_{j}A_{m}\rangle \langle A_{k}\rangle }{\langle A_{i}A_{k}\rangle -
\langle A_{i}\rangle \langle A_{k}\rangle }+ \delta_{ij} \ln Z(t_{0}) \right. \nonumber \\
\left.  -\sum_{m}
\frac{\langle A_{j}A_{m}\rangle -\langle A_{j}\rangle \langle A_{m}\rangle }{\langle A_{i}A_{m}\rangle -
\langle A_{i}A_{m}\rangle }\right] \, . \label{r}
\end{eqnarray}

The values of the correlators $\langle A_{i}\rangle $, $\langle A_{i}A_{k}\rangle $,
$\langle A_{j}A_{k}A_{m}\rangle $ are averaged with $\rho(t)$ and are
$t$-dependent. In deriving (\ref{D}), (\ref{r}) we used the
relations like

$$
\frac{\partial \ln Z(c)}{\partial\langle A\rangle }=\sum_{n}\frac{\partial
\ln Z(c)}{\partial F_{n}(c)}\frac{\partial F_{n}(c)}{\partial
\langle A\rangle }=\sum_{n}\frac{\langle A_{n}\rangle }{\langle AA_{n} \rangle-\langle A\rangle \langle A_{n} \rangle}; \quad
A_{n}(-c)=e^{\textstyle -icL}A_{n};
$$

$$
\frac{\partial F_{m}}{\partial
\langle A_{i}\rangle }=\frac{1}{\langle A_{i}A_{m}\rangle -\langle A_{i}\rangle \langle A_{m}\rangle }.
$$

One can proceed with the expression (\ref{r}) using the relations

$$
\frac{\partial F_{i}}{\partial t}=\sum_{k}\frac{\partial
F_{i}}{\partial \langle A_{k}\rangle }\frac{\partial \langle A_{k}\rangle }{\partial
t}=\sum_{k} {\frac{\partial F_{k}}{\partial
\langle A_{i}\rangle }\frac{\partial \langle A_{k}\rangle }{\partial t};\;\;\frac{\partial
F_{i}}{\partial \langle A_{k}\rangle }=\frac{\partial F_{k}}{\partial
\langle A_{i}\rangle }=\frac{\partial^{2}S}{\partial \langle A_{k}\rangle \partial
\langle A_{i}\rangle }\,;}
$$

$$
\frac{\partial \ln Z}{\partial t}=-\sum_{m} \frac{\partial
F_{m}}{\partial t}\langle A_{m}\rangle .
$$

The solution to (\ref{in}) has the form

\begin{equation}
p_{q}(t-t_{0})=\frac{p_{q}(0)M}{1+p_{q}(0){\displaystyle\int\limits_{t_{0}}^{t}}
{\displaystyle\frac{1}{F_{i}}}r_{2}Mdt} \,. \label{res}
\end{equation}

Since $C_{i}$ and $F_{i}(t_{0})$ does not depend on $t$,

\begin{eqnarray}
M(t)=\exp\left\{-\frac{C_{i}}{F_{i}}(t-t_{0})-\frac{1}{F_{i}}\int_{t_{0}}^{t}
 r_{1}dt\right\}=\nonumber \\
\exp\left\{-\frac{C_{i}}{F_{i}} (t-t_{0})-
\frac{1}{F_{i}}(D(t)-D(t_{0}))\right\} \ , \label{M}
\end{eqnarray}
$M(t_{0})=1$, where $\int_{t_{0}}^{t} r_{1}dt$  is written in
(\ref{D}).

Integrating by parts the second term in
the denominator of (\ref{res}) write it in the following form:

\begin{equation}
p_{q}(t-t_{0})=\frac{p_{q}(0)M(t)}{1-L}; \label{res1}
\end{equation}
$$
L=p_{q}(0)\frac{1}{F_{i}} \left[\left(\int
r_{2}dt\right)_{|t_{0}}-\left(\int r_{2} dt\right)_{|t} M(t)-
\int_{t_{0}}^{t}\left(\int r_{2}dt\right)\left(\frac{C_i}{F_{i}} +
\frac{1}{F_{i}} \frac{dD(t)}{dt}\right)M(t)dt\right],
$$
where
\begin{eqnarray}
\int r_{2}{dt}=\sum_{j} \sum_{m}F_{m}(t_{0})
F_{j}(t_{0})\sum_{k}
\frac{\langle A_{k}A_{j}A_{m}\rangle -\langle A_{j}A_{m}\rangle \langle A_{k}\rangle }{\langle A_{i}A_{k}\rangle -
\langle A_{i}\rangle \langle A_{k}\rangle  }+ \label{res2} \\
F_{i}\ln Z(t_{0})
-\sum_{m} \sum_{j}
 F_{j}(t_{0})\frac{\langle A_{j}A_{m}\rangle -\langle A_{j}\rangle \langle A_{m}\rangle }{\langle A_{i}A_{m}\rangle -
\langle A_{i}\rangle \langle A_{m}\rangle } ; \nonumber
\end{eqnarray}

\begin{eqnarray}
\frac{dD(t)}{dt}=\sum_{m,n}
 \left\{F_{n}(t_{0})\frac{\langle A_{m}A_{n}\rangle -\langle A_{m}\rangle \langle A_{n}\rangle }{\langle A_{m}A_{i}\rangle -
\langle A_{m}\rangle \langle A_{i}\rangle }\left[\frac{d}{dt}\ln (\langle A_{m}A_{n}\rangle - \langle A_{m}\rangle \langle A_{n}\rangle )- \right. \right.
\label{dD}
\\
\left.
-\frac{d}{dt}\ln(\langle A_{m}A_{i}\rangle -
\langle A_{m}\rangle \langle A_{i}\rangle )\right]+F_{m}(c_{1})\frac{\langle A_{m}A_{n}\rangle -
\langle A_{m}\rangle \langle A_{n}\rangle }{\langle A_{i}A_{n}\rangle -\langle A_{i}\rangle \langle A_{n}\rangle }\times \nonumber \\
 \left[\frac{d}{dt}\ln(\langle A_{m}A_{n}\rangle - \langle A_{m}\rangle \langle A_{n}\rangle )-
\frac{d}{dt}\ln(\langle A_{i}A_{n}\rangle -\langle A_{i}\rangle \langle A_{n} \rangle )\right]- \nonumber \\
-F_{m}(c_{1})F_{n}(t_{0})\sum_{k}
\frac{\langle A_{k}A_{m}A_{n}\rangle -\langle A_{m}A_{n}\rangle \langle A_{k}\rangle }{\langle A_{i}A_{k}\rangle -
\langle A_{i}\rangle \langle A_{k}\rangle } \left[\frac{d}{dt}\ln(\langle A_{k}A_{m}A_{n}\rangle -
\langle A_{m}A_{n}\rangle \langle A_{k}\rangle )- \right. \nonumber \\
\left.\left. -\frac{d}{dt}\ln(\langle A_{i}A_{k}\rangle -\langle
A_{i}\rangle \langle A_{k}\rangle )\right]\right\} \,.\nonumber
\end{eqnarray}

The expression for $M(t)$ is given in (\ref{M}), and $p_{q}(0)$ is
determined from the conditions for the norm $\int_{0}^{\infty}p_{q}(u)du=1$
and for the average lifetime $\langle \Gamma\rangle =\int_{0}^{\infty}up_{q}(u)du$.

If one either considers the stationary case or assumes a weak
time dependence in the correlators in (\ref{D})-(\ref{r}),
$D(t)\simeq D(t_{0}), r_{2}\simeq 0$, and (\ref{res})-(\ref{res1})
take on the form

\begin{equation}
p_{q}(t-t_{0})=p_{q}(0)\exp\{-{\frac{C_{i}}{F_{i}}}(t-t_{0})\}
\label{exp}
\end{equation}
with $C_{i}/F_{i}=p_{q}(0)=1/\langle \Gamma\rangle $. In
(\ref{exp}) an exponential distribution for $p_{q}(u)$, used in
\cite{zub74,zub80,zub95} is obtained. However generally the
correlators in (\ref{D})-(\ref{r}) are time dependent. Applying
the full form of (\ref{res1}) for concrete systems it is possible to
state the conditions where the denominator in (\ref{res1}) is
considerably different from unity, hence the lifetime distribution
essentially deviates from the exponential one (\ref{exp}).

The term $L$ in the denominator of (\ref{res1}) is small, since
$p_{q}(0)\approx 1/\langle\Gamma\rangle \ll 1$. In the denominator
of (\ref{res1}) it stands in the combination with \textit{1},
\textit{ $1-L\approx 1$ }. Therefore one can write
$p_{q}=p_{q}(0)M[1+L+L^{2}+\dots ]$ , \ and the distribution
(\ref{res1}) is close to the exponential distribution (\ref{exp})
used in \cite{zub74,zub80,zub95,ram,luz}. This results agrees with
the results of \cite{str}, where the exponential distribution for
the lifetime is shown to be a limiting distribution. But the
expression for $\langle\Gamma\rangle$  is explicitely given in
(\ref{exp}), and in (\ref{D})-(\ref{dD}). It was already pointed
above that the situations where \textit{L} is comparable to 1 can
arise as well.

For the maximum entropy principle with the Shannon measure of the
information entropy the exponential distribution used in
\cite{zub74,zub80,zub95}, is basic. Choosing another measures for
the information entropy (e.g. Tsallis and Renyi measures
\cite{tsal,ren}) changes the function $f_{ME}$ (\ref{dis}), which
yields another forms of the lifetime distributions. For the NSO
method the functions $f_{ME}$ are in the form (\ref{dis}), and the
information entropy is given by the Shannon measure, hence basic
distribution is the exponential one. \

The expressions (\ref{res}), (\ref{res1}) for $p_{q}(u)$ depend on
$F_{i}$, the functions $C_{i}=\sum_{j} C_{ji}F_{j}$,
$r_{k},\;k=1,2$, in (\ref{in})-(\ref{r1}) depend on the index $i$,
the function $p_{q}(u,i)$ depends on $i$ as well. One
can get an $i$-independent function $p_{q}(u)$ by symmetrizing
the distribution, for example, using the operation

$$
p_{q}(u)=\left[\prod_{i=1}^{M} p_{q}(u,i)\right]^{\textstyle \frac{1}{M}}.
$$

Such formulation of maxent principle, as
in \cite{sch}, gives the distribution for the lifetime, related to
the exponential distribution which serves in this case as the base
one. Distributions of other type can be obtained using
some other form of maxent principle.

\section{Another approach to the method of maximum entropy}

Yet another approach to the determination of the type of function of
distribution of lifetime is related to the method of maximal entropy inference ("maxent"),
developed in \cite{stBe} for the determination of the distribution of
superstatistics. We note a formal similarity between the
superstatistics method where the averaging is performed over the
parameter  $\beta$, (for example, the inverse temperature)

$$
p(E)\propto \int_{0}^{\infty}f(\beta)e^{-\beta E}d \beta \,,
$$
and the NSO method where the averaging is performed over the past
life spans $u=t-t_{0}$,

$$
\ln\rho(t)=\int_{0}^{\infty}p_{q}(u)\ln\rho_{q}(t-u, -u)du; \
u=t-t_{0 },
$$
which was already used in \cite{ry07:1}. Therefore the maxent
method of \cite{stBe} can be applied for determining the function
${p_{q}(u)}$. The analogy here is not merely formal. The principal
assumption of \cite{stBe}, that is the separation of the time
scales is essential for the NSO method as well
\cite{zub74,zub80,zub95}. In the approach of superstatistics
\cite{Be3,Be5,Be4} the system is split into cells and local
fluctuations of the value $\beta$  are considered; the
fluctuations of the value $u=t-t_{0}$ affect the complete system.

Closely related is also the research of Crooks \cite{cro}. He
studies general non-equilibrium systems, without assuming that the
system can be divided into different cells that are at local
equilibrium. Crooks claims that instead of trying to obtain the
probability distribution of the entire non-equilibrium system, one
has to try to estimate the ``metaprobability,'' the probability of
the microstate distribution. Crooks also uses the
maximum entropy principle but sets in (\ref{be}) $\lambda_{3}=0$.
The main difference is that Crooks does not assume a local
equilibrium in the cells, hence his approach, though being an
interesting theoretical construction, does not give a
straightforward physical interpretation to the fluctuating
parameter  $\beta$. However such an approach can be applied to the
fluctuations of the value $u=t-t_{0}$. In the approach of
\cite{stBe} one obtains a local fluctuating temperature that
coincides with the thermodynamic temperature and which can in
principle be measured. The work of Crooks is used by Naudts
\cite{nau} to describe equilibrium systems. The author shows that
some well-known results of the equilibrium statistical mechanics
can be reformulated in a very general context with the use of the
concepts introduced in \cite{stBe,Be5,cro}.

In \cite{stBe} following expression is obtained for the
distribution function of the superstatistics \
$f(\beta;\lambda_{i})$

$$
f(\beta;\lambda_i)= \frac{Z(\beta)^{-\lambda_1/V}}{Z(\lambda_i)}
\exp(-\beta \lambda_2 \frac{E(\beta)}{V}-\lambda_3 g(\beta)),
$$
where $\lambda_{i}$ are Lagrange multipliers, $V$ being an
arbitrary constant (taking out a common factor out of the
definition of $\lambda_{1}$ and $\lambda_{2}$ will turn out to be
useful in the following). Using the well-known formula
$S_{\beta}=\ln Z(\beta)+\beta E(\beta)$ with $Z(\lambda_{i})$ being a
normalization constant that is fixed by the condition
$\langle 1\rangle_{\beta}=1$.

The same approach with said limitations used for the function
$p_{q}(u)$, yields

\begin{equation}
p_{q}(u;\lambda_{i})=\frac{Z(t-u)^{-\lambda_{1}/V}}{Z(\lambda_{i})}
\exp(-\lambda_{2}\frac{\sum_{m}
F_{m}(t-u)\langle A_{m}\rangle }{V}-\lambda_{3}{g(u))} ,
\label{be}
\end{equation}
where $g(u)$ is an arbitrary function of $u$, which is determined
by the physical peculiarities of the behaviour of the system in
one or another period of its history. Expression (\ref{be}) is obtained
by the optimization of the entropy $S(\lambda_{i})=\int
p_{q}(u)\ln p_{q}(u)du$ with the constraints for entropy
$$
\int p_{q}(u)S(u)du=\int p_{q}(u)\int
\rho_{q}(t-u,-u)\ln\rho_{q}(t-u,-u)dz du=
$$
$$
\int p_{q}(u)[-\sum_{m}F_{m}(t-u)\langle A_{m}\rangle -\ln Z(t-u)]du
$$
and parameters  $\int p_{q}(u)\sum_{m} F_{m}(t-u)\langle
A_{m}\rangle du$. Similarly to \cite{stBe}, one can set the
functions $g(u)$ in a different fashion. For example, for $g(u)=u,
\lambda_{1}=\lambda_{2}=0, \lambda_{3}=1/\langle \Gamma\rangle $,
where $\langle \Gamma\rangle $ is the average span of past life of
a system (till the present time moment), we get the exponential
distribution used in \cite{zub74,zub80,zub95}. Setting $g(u)=\ln u$
with appropriate corresponding values of $\lambda$ one gets the power-like
distribution for ${p_{q}(u)}$; setting $g(u)=(\ln u)^{2}$ with
corresponding $\lambda$ results in the log-normal distribution.
Thus setting the function $g(u)$ and $\lambda$ accordingly it is
possible to obtain various distributions for the lifetime
considered, for example, in \cite{cox}.

It is possible to examine more difficult
cases when the behaviour of the system changes at different stages of its
evolution, when, for example the function $g(u)=\ln u$ yields the
power-like function  $p_{q}(u)$  at $u < c$, and $g(u)$ gives
an exponential shape of $p_{q}(u)$  at $u > c$.

\section{Conclusion}

For the determination of the lifetime distribution in the NSO method the
method of maximum entropy principle as in \cite{sch} is used. The
obtained distribution is close to exponential $p_{qzub}(u)$
(\ref{zub}), but does not coincide with it. It is possible to find
conditions at which this difference is essential. Using
other variants of the maximum entropy principle, as in
\cite{stBe}, it is possible to obtain other distributions
except exponential one, in particular, power-like and log-normal distributions,
transitory ones between them, as well as distributions of other classes.

In the interpretation of \cite{zub80} it is the random value
$t_{0}$  in  $u=t-t_{0}$  that fluctuates. In \cite{zub80} the
limiting transition is performed for the parameter $\varepsilon,
\varepsilon\rightarrow 0$ in the exponential distribution
$p_{q}(u)=\varepsilon \exp\{-\varepsilon u\}$ after the
thermodynamic limiting transition. In the interpretation of
\cite{ry01} it corresponds to the average lifetime of a system
tending to infinity: $\langle \Gamma\rangle =\langle t-t_{0}\rangle
=1/\varepsilon\rightarrow\infty$. But the average intervals
between successive random jumps grow infinitely, getting larger
than the lifetime of a system. Therefore the source term in the
Liouville equation turns to \textit{0}. If however the
distribution ${p_{q}(u)}$ changes over the interval of the
lifetime, the influence of the environment which caused this
change, remains within the life span even if the lifetime tends to
infinity \cite{ry09}.

There are numerous experimental
confirmations for such change of the lifetime distribution
$p_{q}(u)$ over the interval of the system lifetime.
The examples thereto are the transition to chaos and the
transition from laminar to turbulent flow which are accompanied by
the change of the distribution of  $p_{q}(u)$. In
\cite{in,man} the passage from Gaussian to non-Gaussian
behaviour of the distribution of the first-passage time for some
time moment is demonstrated. Besides the real systems possess
finite sizes and finite lifetime. Therefore influence of
surroundings on them is always present.

Slow change of the function
$g(u)=u=t-t_{0}$ corresponds in the interpretation of \cite{zub80} to the
slow change of the
random value $t_{0}$ on a temporal scale. Accordingly slow
change of the function $g(u)=\ln u$ corresponds to the slow change of $\ln(t-t_{0})$.
Setting other
functions $g(u)$, for example,  $g(u)=(\ln u)^{2}$ and
so on is explained on the same footing.

In the present work by means of two variants of the method of maximum
entropy we obtained the expressions for the distribution of the
lifetime value. It is noted that the choice of the form of the
distribution function for the lifetime value can affect the
non-equilibrium behavior of a system even after performing the
thermodynamic limiting transition.


\section*{References}


\begin{thebibliography}{99}

\bibitem{zub74} Zubarev D N 1974 \emph{Nonequilibrium statistical
thermodynamics} (New York: Plenum-Consultants Bureau)

\bibitem{zub80} Zubarev D N 1980 in: Reviews of Science and Technology: \emph{Modern Problems of
Mathematics}, \textbf{15}, 131-226, (in Russian) ed. by R.B.
Gamkrelidze (Moscow: Nauka)  [English Transl.: J. Soviet Math.
\textbf{16} (1981) 1509]

\bibitem{zub95} Zubarev D N, Morozov V and R\"opke G 1996
\textit{Statistical mechanics of nonequilibrium processes} vol 1
Basic Concepts, Kinetic Theory (Berlin: Akad. Verl.)

\bibitem{Rop} Der R, R\"opke G 1983 \textit{Phys. Lett.A} \textbf{95A} 347

\bibitem{Der} Der R 1985 \textit{PhysicaA} \textbf{132A} 74

\bibitem{alg} Algarte A C, Vasconcellos A R, Luzzi R and Sampaio A J 1985
\textit{Revista Brasileira de Flsica} \textbf{15} 106

\bibitem{ram} Ramos J G, Vasconcellos A R and Luzzi R 1999 \textit{Fortschr.
Phys./Progr. Phys}. \textbf{47} 937

\bibitem{luz} Luzzi R, Vasconcellos A R and Ramos J G 2000
\textit{Statistical Foundations of Irreversible Thermodynamics}
(Stuttgart: Teubner-BertelsmannSpringer)

\bibitem{kir} Kirkwood J G, 1946
\textit{J.Chem.Phys}. \textbf{14} 180; \textit{J.Chem.Phys}.
\textbf{15 } 72.

\bibitem{ry01} Ryazanov V V 2001
\textit{Fortschritte der Phusik/Progress of Physics}
\textbf{49}~885

\bibitem{ry07} Ryazanov V V 2007 \textit{Low Temperature Physics} \textbf{33}
1049

\bibitem{ry09} Ryazanov V V 2009 cond-mat:0902.1454

\bibitem{str}Stratonovich R L 1967 \textit{The
elected questions of the fluctuations theory in a radio
engineering}  (New York: Gordon and Breach)

\bibitem{mor} Morozov V G and R\"opke G, 1998 \textit{Condensed
Matter Physics} \textbf{1~673}

\bibitem{fel} Feller W 1971 \textit{An Introduction to Probability Theory and its
Applications} vol.2 (New York: J.Wiley)

\bibitem{pri} Prigogine I 1980 \textit{From Being to Becoming} (San Francisco:
Freeman)

\bibitem{gre} Green M 1952 \textit{J. Chem. Phys}. \textbf{20} 1281; 1954 ibid.
\textbf{22} 398

\bibitem{mori} Mori H, Oppenheim I and Ross J 1962 in \textit{Studies in
Statistical Mechanics} I, edited by J. de Boer and G.E. Uhlenbeck,
(Amsterdam: North-Holland) pp. 217-298

\bibitem{sch} Sch\"onfeldt J-H, Jiminez N, Plastino A R, Plastino A and Casas M
2007 \textit{Physica A} \textbf{374} 573

\bibitem{jay} Jaynes E T 1957
Phys. Rev. \textbf{106} { 620 }

\bibitem{jay1} Jaynes E T 1957
\textit{Phys. Rev.} \textbf{108}{ 171}

\bibitem{pel} Peletminskii S V and Yatsenko
A A 1968 \textit{Soviet Phys
JETP}\textbf{\textit{26}} {~773;
1967 } \textit{Zh. Eksp. Teor.
Fiz.} \textbf{53}, 1327

\bibitem{akh} Akhiezer A I and Peletminskii
S V 1981 \textit{Methods of statistical
physics}  (Oxford: Pergamon)

\bibitem{tsal} Tsallis C 1988 \textit{J. Stat.Phys}. \textbf{52 }479;
http://tsallis.cat.cbpf.br/biblio.htm .

\bibitem{ren}R\'enyi A 1961
\textstyleHTML{\textit{Proceedings of
the 4th Berkeley Symposium on Mathematics, Statistics and
Probability }}\textstyleHTML{547}

\bibitem{stBe} Van der Straeten E and Beck C 2008
\textit{Phys. Rev. E} \textbf{78} 051101

\bibitem{ry07:1} Ryazanov V V 2007 cond-mat:07101764

\bibitem{Be3} Beck C and Cohen E G D 2003  \textit{Physica A} \textbf{322} 267

\bibitem{Be5} Beck C and Cohen E G D and Swinney H L 2005 \textit{Phys. Rev. E}
\textbf{72} 056133

\bibitem{Be4} Beck C and Cohen E G D 2004 \textit{Physica A} \textbf{344}
393

\bibitem{cro} Crooks G E 2007 \textit{Phys. Rev. E} \textbf{75} 041119

\bibitem{nau} Naudts J 2007 \textit{AIP Conf. Proc. No. 965} (NY: AIP, Melville)
p. 84

\bibitem{cox} Cox D R and Oakes D  1984
\textit{Analysis of Survival Data}  (London, New York: Chapman and Hall)

\bibitem{in} Inoue Jun-ichi and Sazuka N 2007 \textit{Physical Review E}
\textbf{76} 021111(9)

\bibitem{man} Mantegna R N and Stanley H E 1994 \textit{Phys. Rev. Lett},
\textbf{73} 2946

\end {thebibliography}
\end{document}